\documentclass[aps,prl,twocolumn,superscriptaddress,amsfont,amssymb,amsmath,nofootinbib,showpacs,balancelastpage]{revtex4-1}
\usepackage{graphicx,longtable,mathrsfs,color}
\usepackage[usenames,dvipsnames]{xcolor} 
\usepackage{amssymb,amsmath,mathtools,mathrsfs} 
\usepackage{epsfig,subfigure} 
\usepackage{booktabs,longtable} 
\usepackage{exscale,relsize} 
\usepackage[normalem]{ulem} 
\usepackage{enumerate}

\newcommand{\Od}{{\mathcal O}}
\begin{document}

\title{A few cosmological implications of tensor nonlocalities }

\author{Pedro G. Ferreira}
\email{p.ferreira1@physics.ox.ac.uk}
\affiliation{Astrophysics, University of Oxford, DWB, Keble Road, Oxford OX1 3RH, UK}
\author{Antonio L. Maroto}
\email{maroto@ucm.es}
\affiliation{Departamento de F\'{\i}sica Te\'orica, Universidad Complutense de Madrid, 28040 
Madrid, Spain}
\date{Received \today; published -- 00, 0000}

\begin{abstract}
We consider nonlocal gravity theories that include tensor nonlocalities. We show that in the cosmological 
context, the tensor nonlocalities, unlike scalar ones, generically give rise to growing modes. An explicit example with quadratic curvature terms 
is studied in detail. Possible consequences for recent nonlocal cosmological models proposed
in the literature are also discussed.
\end{abstract}


\maketitle

\section{Introduction}
The possibility that General Relativity (GR) may be modified on different length scales is an area of active research \citep{Clifton:2011jh}. The overwhelming majority of theories proposed to replace  GR involve local functions of the degrees of freedom. Some of the most popular proposals work purely with the metric tensor, constructing functions of the curvature tensors and scalars \cite{Sotiriou:2008rp}. Others couple the metric with other new gravitational degrees of freedom such as scalars \cite{Brans:1961sx,Horndeski:1974wa}, vectors \cite{2007PhRvD..75d4017Z,2009PhRvD..80f3512B}  and tensors \cite{2009PhRvD..79f3511B,2010PhRvL.105a1101B,2011PhRvL.106w1101D,2012JHEP...02..126H} which can arise in a variety of setting. Very few proposals consider the possibility that the gravitational action might be fundamentally {\it nonlocal}.

In some sense, the minimalist approach to modifying GR is to look at the effective action for gravity that arises from one loop correction from gauge and matter fields \cite{Barvinsky:1987uw, Barvinsky:1990up, Barvinsky:1990uq,Barvinsky:1995jv}. While, at first sight, nonlocal interactions may seem to be a step too far, they seem to consistently crop up and when they do, they can have far-reaching consequences. So, for example, the feted trace anomaly which was uncovered in the 1970s \cite{Capper:1974ic} played an instrumental role in the first model of inflation constructed by Starobinsky \cite{Starobinsky:1980te}. Surprisingly, it is this model that seems best to fit the current measurements of large scale structure coming out of the analysis of the Planck satellite data \citep{Ade:2013zuv}. The role of the trace anomaly in the early universe was also highlighted in \cite{Dolgov:1993vg} where it was shown that it could be used to generate large scale seed magnetic fields. While in detail, the proposal is problematic (it needs many fermion families and  large gauge groups) it is still one of the frontrunners in explaining how these field formed in the early universe.
Over the years, a number of proposals for the origin, and effect of, nonlocal terms in the gravitational action have been put forward \cite{Wetterich:1987fm,Banks:1988je,Tsamis:1997rk,Espriu:2005qn,Hamber:2005dw}. 

More recently, Deser and Woodard \cite{Deser:2007jk} have embraced the possibility of nonlocal terms and proposed that the gravitational action could take the form
\begin{eqnarray}
S_{DW}=-\frac{1}{16\pi G}\int d^4\sqrt{g}R\left(1+f\left(\frac{R}{\Box}\right)\right) \label{DW}
\end{eqnarray}
Such an action has two appealing properties. For a start, the argument, $R/\Box$, is dimensionless which
means that we expect $f(x)$ to be solely constructed in terms of dimensionless constants of order one- in some sense it is "natural". But, as the authors showed, the fact that it is divided by $\Box$ means that the
modification become large for large length scales and, more interestingly, for long {\it time} scales. 
Given that $R=0$ in pure radiation, they show that it is possible to construct a model which affects late time expansion yet is triggered by the radiation-matter transition \cite{Deser:2013uya}. A few authors have explored alternative local formulations constructed with additional scalar degrees 
of freedom \cite{Nojiri:2007uq}, \cite{Jhingan:2008ym}, \cite{Koivisto:2008xfa}, \cite{Nojiri:2010pw},  \cite{Elizalde:2013dlt}.

It has also been argued that nonlocal terms in the gravitational action can be brought in to generate a graviton mass without the usual pathologies that arise in standard (but not modern) massive gravity \cite{Jaccard:2013gla}. The authors showed that such a modification did not lead to a VDW discontinuity, did not have ghosts \cite{Biswas:2011ar} and, conveniently, led to dark energy like behaviour at late time \cite{Maggiore:2013mea}. The phenomenon is akin to that of "degravitation" which was advocated in \cite{ArkaniHamed:2002fu}.

Nonlocal terms may also introduce modifications to the effective action for gauge fields \cite{Drummond:1979pp,Shore:1995fz,Riegert:1984kt}. Indeed, the complete effective action for gravity and gauge fields \cite{Barvinsky:1987uw, Barvinsky:1990up, Barvinsky:1990uq,Barvinsky:1995jv} clearly shows how an abundance of nonlocal terms of various forms and guises can emerge. It seems then that a sensible approach is to be ecumenical and study a wide range of nonlocal terms; a particularly simple setting is to find -as was done in \cite{Deser:2007jk}- how they effect the background expansion.

In this paper we extend the work of previous authors and look at general quadratic terms in the curvature tensor with a specific form of nonlocality (of the form "$1/\Box$"). We determine the modified field equations and how the standard cosmological solutions is affected. We also look at how nonlocal corrections affect the photon dispersion relations, extending the work of \cite{Shore:1995fz}. In general we find that a growing mode generates a large instability which rapidly dominates at late time. This means that great care must be had when studying these theories- such instabilities seem generic. Nevertheless, we argue that these terms should be studied in more detail and discuss possible limitations to our results.

\section{A recap: the origin of nonlocal terms and the case of $R/\Box$}
The one-loop effective action obtained by integrating out quantum fields 
propagating on gravitational and/or gauge backgrounds is generically a nonlocal 
object which contains all the relevant information about the  semiclassical evolution
of the background fields.  Different techniques to calculate it have been proposed in the 
literature, the most popular one being the so called Schwinger-De Witt
\cite{Schwinger:1951nm,dewitt}
proper time representation. This method  provides an asymptotic expansion in inverse powers
of a mass parameter whose coefficients themselves involve powers of curvatures and their derivatives.
This expansion is covariant but local and can be typically applied when the 
variation scale of the background fields $L$ is much larger than the Compton 
wavelength of the quantum field which is integrated out, i.e for $L\gg 1/m$.
This means that for massive fields, the effects of the new terms in the effective
action are negligibly small on cosmological contexts. Even for massless fields 
the Schwinger-De Witt expansion only contains information about the local
divergences but not on the finite nonlocal terms  
which could be relevant on large scales. 

In order to overcome these difficulties some techniques have been proposed such as  the 
partial resummation of the Schwinger-De Witt expansion 
\cite{resum} obtaining part of the nonlocal structure.
On the other hand, the so called covariant perturbation theory \cite{Barvinsky:1987uw, Barvinsky:1990up, Barvinsky:1990uq} which is based on the fact that 
in asymptotically flat backgrounds, the one-loop effective action is an analytic functional
of the curvatures, allows to expand the effective action in a nonlocal basis of curvature invariants
up to any given order $N$.  
These methods show that for generic quadratic theories with inverse propagator:
\begin{eqnarray}
H=g^{\mu\nu}\nabla_\mu\nabla_\nu+\left(\hat P-\frac{1}{6}R\right) \nonumber
\end{eqnarray}
(where $\hat P$ is the potential matrix of the field) the effective action contains  generic nonlocal terms  in the form:
\begin{eqnarray}
W_{NL}&=&\int d^4x \sqrt{g}\left(\sum_i \gamma_i (-\Box_2){\cal R}_1{\cal R}_2\right.\nonumber \\
&+&\left.
\sum_i \Gamma_i (-\Box_1,-\Box_2,-\Box_3){\cal R}_1{\cal R}_2{\cal R}_3+\Od({\cal R}^4)\right) \nonumber
\end{eqnarray}	
where the $i$ indices run over the dimension of
the corresponding curvature basis.  The generic curvatures ${\cal R}=(R_{\mu\nu}, \hat{\cal R}_{\mu\nu},\hat P)$
correspond to either the Ricci tensor $R_{\mu\nu}$, the curvature associated to 
the arbitrary connection $ \hat{\cal R}_{\mu\nu}=[\nabla_\mu,\nabla_\nu]$
or  $\hat P$. The nonlocal form factors $\gamma_i$  and $\Gamma_i$
typically contain $\log(\Box)$ or inverse powers of $\Box$
among other terms. 

This kind of curvature expansion, despite the fact of being able to capture the 
essential nonlocal behaviour and reproduce quantum anomalies from the effective action, 
are valid only when: 
\begin{eqnarray}
\nabla \nabla {\cal R}\gg {\cal R}{\cal R} \label{range}
\end{eqnarray}
which, as we will show below, is not necessarily satisfied in cosmological contexts.
The possibility of constructing low-energy expansions i.e. expansions valid in the 
opposite limit $\nabla \nabla {\cal R}\ll {\cal R}{\cal R}$ has been also explored \cite{Avramidi:1994zp},
\cite{Avramidi:2009tn} 
although the results in this case are very limited.

Nonlocal terms involving curvature tensors are also present in the nonlocal braneworld action
of the Randall-Sundrum model considered in \cite{Barvinsky:2001tm,Barvinsky:2002kh}.

Deser and Woodard \cite{Deser:2007jk} have embraced the presence of these nonlocal terms to
construct a nonlocal theory that leads to accelerated expansion. Their idea is based on a few
simple, but key, observations. Consider Equation (\ref{DW}) and take $R$ from a standard
cosmology with either radiation or dust. We have that $R=0$ during the radiation dominated phase but
transits to $R=4/(3t^2)$, where $t$ is physical time. If we take the simplest case of $f(x)\simeq x$ we need to solve for $\psi=R/\Box$, i.e. 
\begin{eqnarray}
\Box\psi={\ddot \psi}+3\frac{\dot a}{a}{\dot \psi}= R\label{scalardal}
\end{eqnarray}
with the appropriate boundary condition  we can easily integrate to find a logarithmically growing mode of the form $\ln(t/t_{\rm eq})$ where $t_{\rm eq}$ is the time of equality. As a result it is possible to get moderate
but non-negligible effects on cosmological time scales. In this case, it is radiation-matter equality that can trigger the onset of what would be interpreted as dark energy domination.
In its simplest incarnation, the $R/\Box$ does not do the job but the authors have shown that it is possible that a suitably retrofitted $f(x)$ can fit the observed expansion rate of the universe \cite{Deffayet:2009ca}. Furthermore it passes the usual astrophysical constraints on GR as well as being free of pathological degrees of freedom such as ghosts \cite{Park:2012cp,Deser:2013uya}. 

While the proposal of \cite{Deser:2007jk} opens up an interesting avenue in gravitational physics, it makes sense to look at it in the wider context that motivates it. In particular, terms in $R/\Box$ do not generally appear in isolation, as discussed above, and it is probable that other terms of a tensorial nature may contribute to, or even dominate, the gravitational dynamics. In the next section we explore the properties of these terms, focusing on a nonlocal Ricci tensor coupling.
\section{An $R^{\mu\nu}\frac{1}{\Box}R_{\mu\nu}$ term}
Let us consider the following nonlocal action for the gravitational sector:
\begin{eqnarray}
S_G=\int d^4x \sqrt{g}\left(-\frac{R}{16\pi G}+ M^2 R^{\mu\nu}\frac{1}{\Box}R_{\mu\nu}\right) \nonumber
\end{eqnarray}
While this is a very specific choice, we believe it illustrates the key features that arise in nonlocal tensorial theories. Varying the action with respect to the metric tensor, we get:
\begin{eqnarray}
\delta S&=&\int d^4x \sqrt{g}\delta g^{\alpha\beta}\left(\frac{1}{16\pi G}G_{\alpha\beta}
+M^2\left(-\frac{1}{2}g_{\alpha\beta}R^{\mu\nu}H_{\mu\nu}\right.\right.\nonumber \\
 &+&2R^\mu_{\;\beta}H_{\mu\alpha}+
\Box H_{\alpha\beta}+g_{\alpha\beta}\nabla_\mu\nabla_\nu H^{\mu\nu}-2\nabla_\mu\nabla_\beta H^{\mu}_{\;\alpha}\nonumber \\
&+&\nabla_\alpha H^{\mu\nu}\nabla_\beta H_{\mu\nu}-\frac{1}{2}\nabla_\lambda H^{\mu\nu}
\nabla^\lambda H_{\mu\nu}g_{\alpha\beta}
\nonumber \\
&-&\frac{1}{2}H^{\mu\nu}\Box H_{\mu\nu} g_{\alpha\beta}+ 2H^{\mu\nu}\nabla_\nu\nabla_\alpha H_{\mu \beta}\nonumber\\
&+& 2\nabla_\mu H^{\mu\nu}\nabla_\alpha
H_{\beta\nu}-2\nabla_\nu H_\beta^{\;\mu}\nabla_\alpha H^{\nu}_{\;\mu}\nonumber \\
&-&\left. \left. 2 H_{\beta}^{\;\mu}\nabla_\nu\nabla_\alpha H^\nu_{\;\mu}
\right)\right) \nonumber
\end{eqnarray}
where we have defined $H_{\alpha\beta}$ from $\Box H_{\alpha\beta}=R_{\alpha\beta}$.
Unlike previous works \cite{Koshelev:2013ida}, the nonlocal term is not an analytic function of
$\Box$ and therefore, in the variation procedure we have followed the definitions in 
\cite{Soussa:2003vv}  in order
to correctly enforce causality with two "formal" partial integrations. Thus, we have used:
\begin{eqnarray}
&&\int d^4x\sqrt{g} R^{\mu\nu}\delta\left(\frac{1}{\Box}R_{\mu\nu}\right)=\nonumber \\
&&\int d^4x\sqrt{g} R^{\mu\nu}\left(-\frac{1}{\Box}(\delta\Box)\left(\frac{1}{\Box}R_{\mu\nu}\right)+\frac{1}{\Box}\delta R_{\mu\nu}\right)=\nonumber \\
&&\int d^4x\sqrt{g} \left(\frac{1}{\Box}R^{\mu\nu}\right)\left(-(\delta\Box)\left(\frac{1}{\Box}R_{\mu\nu}\right)+\delta R_{\mu\nu}\right) \nonumber
\end{eqnarray}

We can rewrite the Einstein field equations in the suggestive form:
\begin{eqnarray}
R_{\alpha\beta}-\frac{1}{2}g_{\alpha\beta}R=8\pi G(T_{\alpha\beta}+T^{NL}_{\alpha\beta}) \nonumber
\end{eqnarray}
where:
\begin{eqnarray}
T^{NL}_{\alpha\beta}&=&-2M^2\left(-g_{\alpha\beta}R^{\mu\nu}H_{\mu\nu}
+R^\mu_{\;\beta}H_{\mu\alpha}+R^\mu_{\;\alpha}H_{\mu\beta}\right.\nonumber \\
 &+&
R_{\alpha\beta}+g_{\alpha\beta}\nabla_\mu\nabla_\nu H^{\mu\nu}
-\nabla_\mu\nabla_{\beta} H^{\mu}_{\;\alpha}-\nabla_\mu\nabla_{\alpha} H^{\mu}_{\;\beta}\nonumber \\
&-&\frac{1}{2}\nabla_\lambda H^{\mu\nu}
\nabla^\lambda H_{\mu\nu}g_{\alpha\beta}+\nabla_\alpha H^{\mu\nu}\nabla_\beta H_{\mu\nu}
\nonumber \\
&+& H^{\mu\nu}\nabla_\nu\nabla_{\alpha} H_{\mu \beta}+ H^{\mu\nu}\nabla_\nu\nabla_{\beta} H_{\mu \alpha}
+
\nabla_\mu H^{\mu\nu}\nabla_{\alpha}
H_{\beta\nu}\nonumber\\
&+&\nabla_\mu H^{\mu\nu}\nabla_{\beta}
H_{\alpha\nu} -\nabla_\nu H_{\beta}^{\;\mu}\nabla_{\alpha } H^{\nu}_{\;\mu}
-\nabla_\nu H_{\alpha}^{\;\mu}\nabla_{\beta} H^{\nu}_{\;\mu}
\nonumber \\
&-&  \left. H_{\beta}^{\;\mu}\nabla_\nu\nabla_{\alpha } H^\nu_{\;\mu}
-  H_{\alpha}^{\;\mu}\nabla_\nu\nabla_{\beta} H^\nu_{\;\mu}\right) \nonumber
\end{eqnarray}
where $T^{\mu\nu}_{NL ; \mu}=0$ has been checked explicitly on the solutions.

Let us now try and solve $H_{\alpha\beta}$. We will mimic \cite{Deser:2007jk} and
assume that the nonlocal terms are a perturbation on a general relativistic background. We will be 
working with the conformal metric $g_{\alpha\beta}=a^2(\eta)(d\eta^2-d{\vec r}^2)$ and, given the
symmetries of the problem, we assume
\begin{eqnarray}
H^\mu_{\;\;\nu}=\mbox{diag}(H_t,-H_s,-H_s,-H_s) \nonumber
\end{eqnarray}
which satisfy 
\begin{eqnarray}
H_t''&+&2{\cal H}H_t'-6{\cal H}^2(H_t+H_s)=3\left({\cal H}^2-\frac{a''}{a}\right) \nonumber \\
H_s''&+&2{\cal H}H_s'-2{\cal H}^2(H_t+H_s)=\left({\cal H}^2+\frac{a''}{a}\right) \nonumber
\end{eqnarray}
The system can be diagonalized if we define two new functions, 
$S=H_t+H_s$ and $D=H_t-3H_s$ and we obtain
\begin{eqnarray}
S''&+&2{\cal H}S'-8{\cal H}^2S=4{\cal H}^2-2\frac{a''}{a} \label{S}\\
D''&+&2{\cal H}D'=-6\frac{a''}{a} \label{D}
\end{eqnarray}
Equations (\ref{S}) and (\ref{D}) are equivalent to those found in \cite{Modesto:2013jea} and present a key feature of tensor nonlocalities. From  (\ref{scalardal}) we see that the D'Alembertian of a scalar field consists solely of derivative terms (recall that the $\Box\psi=\frac{1}{\sqrt{g}}\partial_\mu(\sqrt{g}g^{\mu\nu}\partial_\nu \psi)$). This is not the case for tensor fields where there will be non-derivative terms, schematically of the form
$\sim \Gamma \Gamma H$, where $\Gamma$ is the connection coefficient. Depending on the sign of these terms, they lead to oscillatory or, as we see in (\ref{S}), a growing mode. As we shall see,
this means that the tensor nonlocalities will dominate the dynamics. Notice also that the $D$ mode
which corresponds to the trace of $H^{\mu}_{\;\nu}$ does not exhibit growing solutions
of the homogeneous equation.

To illustrate the point we have just made, we now solve for $S$ and $D$ for particular choices of backgrounds. At late times we can
assume matter domination with $a\propto \eta^2$ and we then find
\begin{eqnarray}
S&=&C_+\eta^{p_+}+C_-\eta^{p_-}-\frac{3}{8} \nonumber \\
D&=&\frac{C_1}{\eta^3}-4\ln \eta + C_2 \nonumber
\end{eqnarray}
where $p_{\pm}=-\frac{3}{2}\pm \frac{\sqrt{137}}{2}$. If the universe is perfectly matter dominated throughout, we would have that the boundary conditions could be such that $C_+=0$ and, again we would have, at most, a logarithmic growth. But this is not the case: we have to match onto a non-zero solution emerging from the radiation era (unlike for the case of the $R/\Box$ term) which will generally
stimulate a $C_+$ term. This dominant, growing, mode evolves as $S\propto \eta^{4.35}\propto a^{2.18}$ and leads to a $T^{NL}_{\alpha\beta}$ such that $\rho_{NL}\propto\eta^{2p_+ -6}$ 
which corresponds to a fluid with an equation of state 
$w_{NL}=-1.451$. That is, the tensor nonlocality is responsible for a dominant growing modification which
rapidly drowns out any slow, desirable, logarithmic modification.

That there is a non-zero solution emerging from the radiation era is easy to show. Choosing  $a\propto \eta$ one has
\begin{eqnarray}
S&=&C_+\eta^{p_+}+C_-\eta^{p_-}-\frac{1}{2} \nonumber \\
D&=&C_1 + \frac{C_2}{\eta} \nonumber
\end{eqnarray}
where $p_{\pm}=-\frac{1}{2}\pm \frac{\sqrt{33}}{2}$. The best one could hope
for is a constant solution but any residual matter or a transition in the effective degrees of freedom will perturb the evolution away and seed a growing mode such that $S\propto \eta^{2.37}\propto a^{2.37}$. We then have
$\rho_{NL}\propto\eta^{2p_+ -4}$ 
so that $w_{NL}=-1.248$.
Finally, for completeness, if we push back to a primordial inflationary regime, we have 
$a\propto -1/\eta$ which leads to
\begin{eqnarray}
S&=&C_+\eta^{p_+}+C_-\eta^{p_-} \nonumber \\
D&=&C_1\eta^3 + 4\ln (-\eta) +C_2 \nonumber
\end{eqnarray}
where $p_{\pm}=\frac{3}{2}\mp \frac{\sqrt{41}}{2}$, and $p_+$ corresponding to the 
growing mode. Again, this is for pure de Sitter and any
small deviation will trigger a growing mode such that  $S\propto (-\eta)^{-1.70}\propto a^{1.70}$
so that $
\rho_{NL}\propto (-\eta)^{2p_+}$ 
and therefore $w_{NL}=-2.134$

In general we find that: 
\begin{eqnarray}
\rho_{NL}\propto \rho_{bk}\vert\eta\vert^{2p_+} \nonumber
\end{eqnarray}
with $\rho_{bk}$ the background energy density driving the expansion. This means that, very rapidly, the growing mode in the tensor nonlocalities comes into play and the modified dynamics dominates the expansion rate of the universe, unless the scale $M$ is 
extremely small. Furthermore, the quadratic terms, which have been generally discarded in the past \cite{Barvinsky:2003kg,Hamber:2005dw,Maggiore:2013mea,Modesto:2013jea} cannot be ignored and dominate the behaviour of $T^{NL}_{\alpha\beta}$.

We have focused on one particular nonlocal, tensorial, term in the gravitational action yet, from the discussion in the previous section and full action presented in \cite{Barvinsky:1987uw, Barvinsky:1990up, Barvinsky:1990uq} we expect other possible terms. So, for example, we might consider nonlocal terms involving the full Riemann tensor of the form
\begin{eqnarray}
\frac{R^{\mu\nu}_{\;\;\;\; \rho\sigma}}{\Box}=H^{\mu\nu}_{\;\;\;\; \rho\sigma} \nonumber
\end{eqnarray}
But these can be rewritten in terms of the Ricci tensor  and scalar; in a Robertson-Walker background, the  Weyl tensor identically vanishes, so that:
\begin{eqnarray}
R^{\mu\nu}_{\;\;\;\; \rho\sigma}&=&g^{\mu}_{\;[\rho}R_{\sigma]}^{\;\nu} -g^{\nu}_{\;[\rho}R_{\sigma]}^{\;\mu}-\frac{1}{3}g^{\mu}_{\;[\rho}g_{\sigma]}^{\;\nu} R \nonumber
\end{eqnarray}
and therefore, using the fact that the metric tensor commutes with $1/\Box$ we have:
\begin{eqnarray}
R_{\mu\nu}^{\;\;\;\; \rho\sigma}\frac{1}{\Box}R^{\mu\nu}_{\;\;\;\; \rho\sigma}=-\frac{1}{3} R\frac{1}{\Box} 
R+2 R_{\mu\nu}\frac{1}{\Box}R^{\mu\nu} \nonumber
\end{eqnarray}
Hence the results that we have found for the $R_{\mu\nu}/\Box$ case can be easily applied here.
Notice also that the same kind of arguments can be applied to terms like $G_{\mu\nu}\frac{1}{\Box}R^{\mu\nu}$ considered in \cite{Barvinsky:2003kg}.

The argument made above is general: anything with an index may lead to a growing mode. So, for example, if consider a nonlocal vector term of the form
\begin{eqnarray}
\frac{V^\mu}{\Box}=H^{\mu} \nonumber
\end{eqnarray}
and restrict ourselves to homogeneity and isotropy such that $H^\mu=(H^0(\eta),0,0,0)$
then we have
\begin{eqnarray}
a^{-4}\left({H^0}''-\left(3{\cal H}^2+\frac{a''}{a}\right)H^0\right)=V^0 \nonumber
\end{eqnarray}
Once again, we find power law solutions for the homogeneous equation: 
\begin{eqnarray}
H^0=C_+ \eta^{p_+}+C_- \eta^{p_-} \nonumber
\end{eqnarray}
with $p_\pm=(1\pm \sqrt{13})/2$ during the radiation era, $p_\pm=(1\pm \sqrt{97})/2$ during the matter era and $p_\pm=(1\pm \sqrt{21})/2$ during the inflationary era.

\section{A nonlocal photon.}

In the previous section we have focused on the gravitational dynamics and how it is affected by nonlocal terms. But a crucial component of cosmology is how we observe the expansion rate- we do this with photons via luminosity and angular diameter distances. 
As shown before, nonlocal terms involving Ricci and gauge curvatures are also expected when
calculating the one-loop effective action and so it makes sense to focus on an abelian gauge theory in a curved space-time and how it may be affected by the simplest tensor nonlocality: 
\begin{eqnarray}
S&=&\int d^4x \sqrt{g}\left(-\frac{1}{4}F^{\mu\nu}F_{\mu\nu}+\alpha\frac{R^\alpha_{\;\;\beta}}{\Box}
F_{\mu\alpha}F^{\mu\beta}\right)\nonumber \\
&=& \int d^4x \sqrt{g}\left(-\frac{1}{4}F^{\mu\nu}F_{\mu\nu}
+\alpha H^\alpha_{\;\;\beta}
F_{\mu\alpha}F^{\mu\beta}
\right) \nonumber
\end{eqnarray}
In such a theory, the modified Maxwell equations are
\begin{eqnarray}
\frac{1}{\sqrt{g}}\partial_\mu (\sqrt{g}F^{\mu\nu})-\frac{2\alpha}{\sqrt{g}}\partial_\mu(\sqrt{g}(H_\alpha^{\;\;\nu}F^{\mu\alpha}-H_\alpha^{\;\;\mu}F^{\nu\alpha}))=0 \nonumber \\
\label{Maxwell}
\end{eqnarray}

Let us now focus on the temporal component of the Faraday tensor (i.e. setting $\nu=0$). We then have 
\begin{eqnarray}
\nabla_\mu F^{\mu 0}-2\alpha\nabla_\mu(H_\alpha^{\;\;0}F^{\mu\alpha}-H_\alpha^{\;\;\mu}F^{0\alpha})=0 \nonumber
\end{eqnarray}
which in a conformal space-time takes the form
\begin{eqnarray}
-\frac{1}{a^4}\partial_i F_{i 0}+\frac{2\alpha}{a^4}\partial_i(H_0^{\;\;0}F_{i 0}-H_j^{\;\;i}F_{0 j})=0 \nonumber
\end{eqnarray}
and greatly simplifies to
\begin{eqnarray}
[1-2\alpha(H_t-H_s)]\partial_i F_{i 0}=0 \nonumber
\end{eqnarray}
where $H_t$ and $H_s$ were defined and calculated in the previous section. With the  Coulomb gauge, $\vec\nabla \cdot\vec A=0$, this leads to $\nabla^2 A_0=0$ and, in the absence of sources we can choose
$A_0=0$.

Let us now consider a wave of the form
$A_\mu=(0,A_i(\eta) e^{i\vec k\vec x})$ such that $\vec k\cdot\vec A=0$. The spatial part
of the Maxwell equations is now
\begin{eqnarray}
\nabla_\mu F^{\mu i}-2\alpha\nabla_\mu(H_\alpha^{\;\;i}F^{\mu\alpha}-H_\alpha^{\;\;\mu}F^{i\alpha})=0 \nonumber
\end{eqnarray}
which leads to
\begin{eqnarray}
A''_i+\frac{1+4\alpha H_s}{1+2\alpha(H_s-H_t)}\vec k^2 A_i+\frac{2\alpha(H'_s-H'_t)}{1+2\alpha(H_s-H_t)} A'_i=0 \nonumber
\end{eqnarray}

We can explore the properties of this equation in light of what we learnt in the previous section.
Taking only the growing mode, we will have that the $S$ part of $H_s$ and $H_t$ will dominate leading to
\begin{eqnarray}
A''_i-\frac{\alpha S'}{1-\alpha S} A'_i+\frac{1+\alpha S}{1-\alpha S}\vec k^2 A_i=0 \nonumber
\end{eqnarray}
Furthermore, asymptotically $\vert S\vert \gg 1$, so that
\begin{eqnarray}
A''_i+\frac{S'}{S} A'_i-k^2 A_i=0 \nonumber
\end{eqnarray}
If, as we saw in the last section, $S\propto \eta^p$ we have that $A_i\propto \eta^{\frac{1-p}{2}}I_q(\vert k\vert\eta)$ where $q^2=(p-1)p/4$ and $I_q(x)$ is a modified Bessel function, i.e. 
 asymptotically the solution diverges exponentially.

To understand the structural effects of the nonlocal term, it is useful 
to solve (\ref{Maxwell}) using the geometric optics approximation \cite{MTW} in 
which the electromagnetic field $A_\mu$ is assumed to behave as a fast evolving phase
times a slowly evolving amplitude, i.e.:
\begin{eqnarray}
A_\mu=(a_\mu+ \epsilon b_\mu+\dots)e^{i\theta /\epsilon} \nonumber
\end{eqnarray}
where the $\epsilon$ parameter controls how rapidly each term evolves. 
Defining $k_\mu=\partial_\mu \theta$ and substituting the expansion in (\ref{Maxwell})
keeping only the leading terms of order $1/\epsilon^2$, we obtain:
\begin{eqnarray}
-k^2 a^\nu+2\alpha H_{\alpha}^{\; \nu}k^2 a^\alpha-2\alpha H_\alpha^{\; \mu}(k_\mu k^\nu a^\alpha-
k_\mu k^\alpha a^\nu)=0\nonumber 
\end{eqnarray}
where $k^2=k_\mu k^\mu$ and we have used the Lorenz gauge condition $\nabla_\mu A^\mu=0$
which implies, to leading order: $k_\mu a^\mu=0$. 

As above we find
\begin{eqnarray}
[1-2\alpha(H_t-H_s)]k^2 a^0=0 \nonumber
\end{eqnarray}
and
\begin{eqnarray}
&-&k^2 a^i-2\alpha H_sk^2 a^i+2\alpha H_t k_0 k^0a^i-2\alpha H_sk_j k^ja^i\nonumber \\
&-&2\alpha H_tk_0 a^0 k^i+2\alpha H_s k_j a^j k^i=0 \nonumber
\end{eqnarray}
whose non-trivial solutions imply $a^0=0$ so that $a^ik_i=0$ and therefore:
\begin{eqnarray}
k_0k^0(1+2\alpha (H_s-H_t))+k_jk^j(1+4\alpha H_s)=0 \label{leading}
\end{eqnarray}
i.e. keeping only the growing mode,  the modified dispersion relation reads:
 \begin{eqnarray}
\omega^2=\frac{1+\alpha S}{1- \alpha S}\;\vec k^2 \nonumber
\end{eqnarray}
with $\omega^2=k_0^2$ and $\vec k^2= k_i  k_i$ 
which agrees with the previous result.

For the $1/\epsilon$ terms, we get:
\begin{eqnarray}
&-&k^2 b^\nu+2H_{\alpha}^{\; \nu}k^2 b^\alpha-2\alpha H_\alpha^{\; \mu}(k_\mu k^\nu b^\alpha-
k_\mu k^\alpha b^\nu)\nonumber \\
&+&2ik^\alpha a^\nu_{;\alpha}+ik^\alpha_{;\alpha}a^\nu
+2i\alpha H_{\alpha\;\; ;\mu}^{\;\nu}(k^\mu a^\alpha- k^\alpha a^\mu)\nonumber \\
&-&4i\alpha H_{\alpha}^{\; \nu}k^\beta a^\alpha_{\; ;\beta}-2i\alpha H_{\alpha}^{\; \nu}k^\beta_{\; ; \beta}a^\alpha+2i\alpha H_{\alpha\;\; ;\mu}^{\; \mu}(k^\nu a^\alpha-k^\alpha a^\nu)\nonumber \\
&+&2i\alpha H_{\alpha}^{\; \mu}
(k^\nu_{\; ; \mu} a^\alpha-k^\alpha_{\; ; \mu} a^\nu)+2i\alpha H_{\alpha}^{\; \mu}
(k^\nu a^\alpha_{\; ; \mu}-k^\alpha a^\nu_{\; ; \mu} )=0\nonumber 
\end{eqnarray}
Using the order $1/\epsilon^2$ equations (\ref{leading}), it is possible to cancel the terms
containing $b^\mu$ in the first line, so that we are left with an equation for $a^\mu$
which describes the propagation of the amplitude and polarization of the 
field. 

Thus, in the Robertson-Walker background assuming also 
homogeneity of $a_\nu$ and $k_\nu$, i.e. 
$\partial_i a_\nu=\partial_i k_\nu$=0 and 
using the expression for $H^\mu_{\; \nu}$, we get:
\begin{eqnarray}
\partial_0\left(k^0 A^2 (1+2\alpha (H_s-H_t))\right)=0 
\label{area}
\end{eqnarray}
where we have written $a_\nu= A f_\nu$,where $A$ is the wave amplitude and
 $f_\nu$ a unit polarization vector with $f^0=0$. Using the previous equation we can
also obtain:
\begin{eqnarray}
\left(1+2\alpha (H_s-H_t)\right)\partial_0 f_i=0 \nonumber
\end{eqnarray}
i.e. the polarization vector remains constant in the evolution, in agreement with the previous results. 
We also find in (\ref{area}) a  modification of the so called area law of ordinary electromagnetism which is directly related to photon number conservation \cite{MTW}. However, in this case, the standard 
interpretation in terms of photons is not appropriate at late times due to the fact that the 
electromagnetic modes will grow exponentially because of the 
tachyon instability.

\section{Discussion}

In this paper, and following on from \cite{Deser:2007jk}, we have explored the role of nonlocal corrections to the dynamics of the Universe. While in \cite{Deser:2007jk} have attempted to flesh out a working model, we have opted to be agnostic and chosen the one loop effective action from \cite{Barvinsky:1987uw, Barvinsky:1990up, Barvinsky:1990uq,Barvinsky:1995jv} as our starting point. As such we have had to address the importance of tensor nonlocalities and have found that, in general, they dominate the evolution of the Universe at 
late times. We have, of course, taken a simplified, perturbative approach and shown that, because of very rapidly growing solution, quadratic terms in the expansion become important. 
The result suggests that higher and higher curvature terms would become more and more important, thus
signaling the breakdown of this kind of nonlocal curvature expansions in the cosmological context.  

A similar result has been obtained in the case of the nonlocal contributions to the electromagnetic action.
In the simplest case with tensor nonlocalities, the growing solution induces a tachyon instability in
the electromagnetic field at late times which again signals the inadequacy of this kind
of truncated nonlocal curvature expansions in cosmology.  
Although in this work we have concentrated on two particular examples, these kinds of instabilities seem to be generic in the case of tensor nonlocalities, in contrast with the  scalar cases in which they are absent. 

Clearly a few things need to be done. For a start it would make sense to undertake a full, self-consistent, analysis of the modified field equations to see if the growth could be suppressed in 
certain particular scenarios in which this kind of effective actions could be
applicable. It is unlikely that this is the case but possible and should be checked.
Furthermore, there are other, possible non-local operators that one might consider. One example would be  $\Box+m^2$, where the mass term might regularise the growing mode. Another example would be the operator that arises with the conformal anomaly, $\Box^2+2\nabla_\mu[R^{\mu\nu}-\frac{1}{3}g^{\mu\nu}]\nabla_\mu$. It is conceivable that such operators might lead to somewhat different behaviour.
 
 More ambitiously, one needs to come up with a more complete effective action for gravity, including more than just the quadratic or cubic terms 
in curvatures that have been considered in this analysis. Going beyond the regime of small enough curvature or large derivatives defined by  (\ref{range}) would  shed new light
on the role of quantum effects on the late time cosmological evolution.


\textit{Acknowledgments.---} We are extremely grateful to R. Woodard for his advice and guidance and to Tessa Baker, Joe Conlon and Johannes Noller for discussions. PGF acknowledges support from Leverhulme, STFC, BIPAC and the Oxford Martin School and the hospitality of the Higgs Centre in Edinburgh while this paper was being completed. ALM acknowledges support from MICINN (Spain) project numbers FIS2011-23000, Consolider-Ingenio MULTIDARK CSD2009-00064, Salvador de Madariaga program and the hospitality of Oxford BIPAC.
\bibliographystyle{apsrev4-1}
\vskip -0.32in
\bibliography{BibliographyNL}
\end{document}